\documentclass[%
 reprint,
superscriptaddress,
 amsmath,amssymb,
 aps,
pre,
]{revtex4-1}

\usepackage{graphicx}
\usepackage{dcolumn}
\usepackage{bm}

\begin{document}


\title{Testing statistical laws in complex systems}

\author{Martin Gerlach}
\affiliation{Department of Chemical and Biological Engineering, Northwestern University, Evanston, IL 60208, USA}

\author{Eduardo G. Altmann}
\affiliation{School of Mathematics and Statistics, University of Sydney, 2006 NSW, Australia}

\date{\today}

\begin{abstract}
  The availability of large datasets requires an improved view on statistical laws in complex systems, such as Zipf's law of word frequencies, the Gutenberg-Richter law of earthquake magnitudes, or scale-free degree distribution in networks. In this paper we discuss how the statistical analysis of these laws are affected by correlations present in the observations, the typical scenario for data from complex systems. We first show how standard maximum-likelihood recipes lead to false rejections of statistical laws in the presence of correlations. 
  We then propose a conservative method (based on shuffling and under-sampling the data) to test statistical laws and find that accounting for correlations leads to smaller rejection rates and larger confidence intervals on estimated parameters.
\end{abstract}

\maketitle


\paragraph*{Introduction}

Statistical regularities collected in the form of ``universal laws'' play a central role in complex systems~\cite{Newman,Mitzenmacher2004,Laws}. Zipf's law of word frequencies~\cite{Zipf}, the Gutenberg-Richter law of earthquake magnitudes~\cite{Sornette}, scale-free degree distributions in networks~\cite{BarabasiAlbert1999}, and inter-event time distributions between bursty events~\cite{Bunde,Barabasi,bursts,burstsC} are prominent examples that triggered entire research lines devoted to explaining the origin and to exploring the consequences of these laws.

Recently, the empirical support of such laws has been heavily questioned. 
The best known example is the case of scale-free degree distribution of networks: after the seminal work of Barabasi and Albert in 1999~\cite{BarabasiAlbert1999}, the early 2000's were marked by findings of power-law distributions in various network datasets, while in the last five years the trend has reversed and it is now common to read that networks with power-law degree distribution are rare~\cite{Khanin,Broido} (see Ref.~\cite{Quanta2018} for a journalistic account). This recent shift in conclusions, which appears in the analysis of Zipf's law in language~\cite{PRX,Laws,Francesc} and also in other areas~\cite{CriticalTruth,CSN}, is partially due to new (larger) datasets but mostly due to the improved statistical methods: least-squared fitting and visual inspection of double-logarithmic plots (used since Zipf) have been replaced by maximum likelihood methods made popular in the influential article by Clauset, Shalizi, and Newman~\cite{CSN}, see Refs.~\cite{Goldstein04,Bauke07,Deluca13,Hanel} for variations. A point often ignored in the interpretations of the recent findings is that these methods rely on two hypotheses:
\begin{itemize}
\item[H1:] The observations $x$ are distributed as $p(x;\vec{\alpha})$, where $\vec{\alpha}$ are parameters, e.g. for a power law
  \begin{equation}\label{eq.powerlaw}
  p(x;\alpha) = C x^{-\alpha}.
\end{equation}

\item[H2:] The empirical observations $x_i, i=1 \ldots N$ are independent (e.g., of $i$ or $x_{i-1}$). 
\end{itemize}
While the statistical laws correspond to H1, the statistical tests rely also on H2 (implicitly assumed, e.g., when the log-likelihood is computed as $\sum_{i=1}^N \log p(x_i)$~\cite{StumpfEPL,Khanin,CSN,Hanel,Sornette}). Complex systems are characterized by strong (temporal and spatial) inter-dependencies~\cite{Eisler.2008} and it is thus not clear whether the recent claims~\cite{Khanin,CriticalTruth,Broido} of violation of the statistical laws arise from systematic deviations of the law itself (H1) or, instead, whether they are due to the well-known fact that observations are not independent (H2).

In this Letter we show that dependencies in the data (violation of H2) have a strong impact on the empirical analysis of statistical laws, leading to rejections even in processes that satisfy the law (H1), and to over-confident selection of models and parameters.  
We then propose an alternative method that distinguishes between H1 and H2, yielding an upper bound on the degree of correlations for which the statistical law is rejected.

\begin{figure*}[bt]
\includegraphics[width=2\columnwidth]{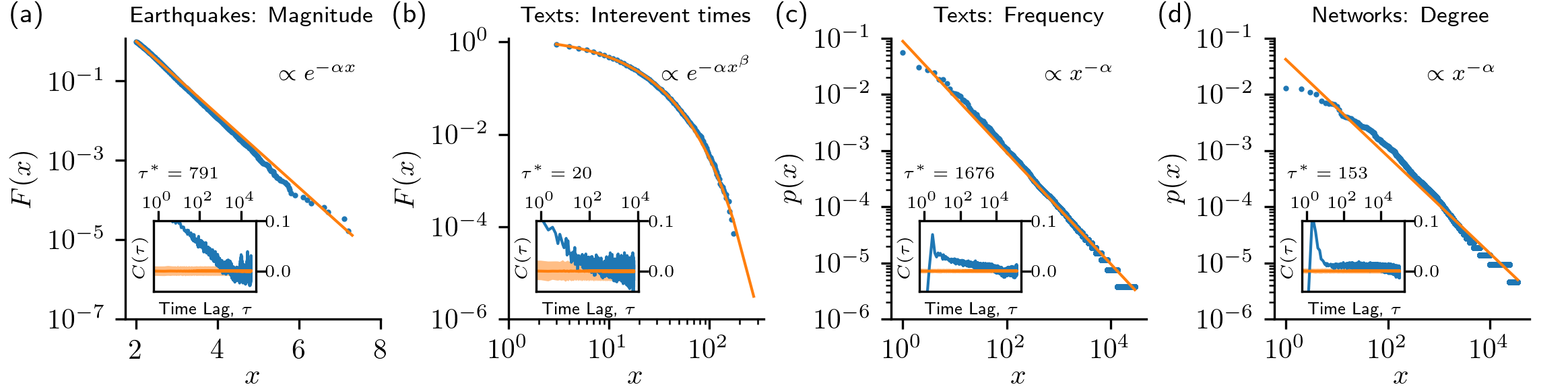}
\caption{
Statistical laws and strong correlations occur simultaneously in complex systems. Main Panels: Distribution $p(x)$ (or its cumulative $F(x)$) of observable $x$ for the data (blue dots) and Maximum-Likelihood fit of different statistical laws (orange, see Supplementary Material, SM, Sec.~I). Insets:  autocorrelation $C(\tau)$ with time lag $\tau$ of the observable $x$ for the original data (blue) and randomized data (orange, average and $1$-/$99$-percentiles over $1000$ realizations); $\tau^*$ indicates the value at which the original and randomized $C(\tau)$ are statistically indistinguishable~\cite{footnote-autocorrelation}.
 \textbf{(a)} Sequence of magnitudes $x$ of earthquakes in Southern California from 1981-2010 \cite{earthquakes} ($N=59,555$ with commonly used threshold $x \geq 2$~\cite{Corral2004}).
  \textbf{(b)} Sequence of interevent times  $x$ (measured in words) of consecutive occurrences of the word ``the'' in the book ``Moby Dick'' obtained from Project Gutenberg~\cite{gutenberg} ($N=14,042$ with threshold $x \geq 3$). 
 \textbf{(c)} Sequence of words (tokens) in the order they appear in the book ``Ulysses'' by James Joyce obtained from Project Gutenberg~\cite{gutenberg}; $x$ the rank of the word (type) in terms of frequency in the whole book  ($N=264,971$ word-tokens,
obtained removing punctuation and non-alphabetic characters).
 \textbf{(d)} Sequence of degrees of nodes from a network; $x$ is the rank of the degree of the node; the sequence $\{x_i\}$ used to compute $C(\tau)$ was obtained applying an edge-sampling method to the complete network (see SM, Sec.~II); the network corresponds to the connections between autonomous systems of the Internet~\cite{network}, $V=34,761$ vertices (nodes) and $E=107,720$ unique edges (in our case $N$ is the number of half-edges and thus $N=2E$).
  }
\label{fig.autocorr}
\end{figure*}

\paragraph*{General setting.} Let $\{x_i\} = x_1,x_2, \ldots, x_N$ be an ordered sequence obtained from a measurement process that asymptotically has a well defined distribution $p(x) = \frac{\# x_i=x}{N}$ as $N \rightarrow \infty$. In observations of dynamical systems (or time series), $x_i$ will typically depend on the observations at previous times so that for all times $\tau$ smaller than some (relaxation) time $\tau^*$ we find  $p(x_i|x_{i-\tau}) \neq p(x)$. Violations of H2 happen also when data  is not measured as a time-series. In the case of Zipf's law of word frequencies, syntax restrict the valid sequences of word tokens, in violation of H2 (both in the rank-frequency and frequency distribution pictures~\cite{Laws,Francesc}). In the case of networks, H2 can be violated because of the generative process or because of the sampling employed to {\it observe} the nodes and links (typically a subsample of an underlying network).
In fact, it has been shown that the degree distribution of networks is sensitive to the sampling procedure~\cite{StumpfPNAS,StumpfPRE,Lee}. Moreover, the hypotheses H1 and H2 of the standard tests for power-law distribution do not build a proper probabilistic network model~\cite{Note1}, are thus not suitable to a rigorous statistical analysis~\cite{Crane}, and the analysis of the degree distribution of networks requires further assumptions about the sampling/generative process. 

More generally, strong correlations are ubiquitous in complex systems~\cite{Eisler.2008} and it is hard to imagine a case for which H2 holds. 
In Fig.~\ref{fig.autocorr} we show how previously proposed statistical laws and correlations appear together in paradigmatic complex systems: the Gutenberg-Richter law for earthquakes (exponential~\cite{Sornette}), interevent times of words (stretched exponential~\cite{Bunde,bursts,burstsC}), Zipf's law for word frequencies (power-law~\cite{Zipf}), and scale-free distribution for the node-degree in networks (power-law~\cite{BarabasiAlbert1999}).
While earthquake events and interevent times naturally occur as time series data, we mapped word frequencies in texts and the network data into ordered sequences $\{x_i\}$ based on a simple sampling process (see caption of Fig.~\ref{fig.autocorr}) in order to illustrate and quantify the violation of H2 in an unified framework.

\paragraph*{Constructed example.} We now show that the traditional methods~\cite{CSN} lead to a rejection of a power-law distribution~(\ref{eq.powerlaw}) even for data which are power-law distributed for $N\rightarrow \infty$.  This is done by building a Markov process~\cite{B1,B2} in which H1 is satisfied but H2 is violated (i.e., $x_i$ depends on $x_{i-1}$ and $p(x) = Cx^{\alpha}$ for $N\rightarrow \infty$, see SM Sec. III).

\begin{figure*}[bt]
\includegraphics[width=2\columnwidth]{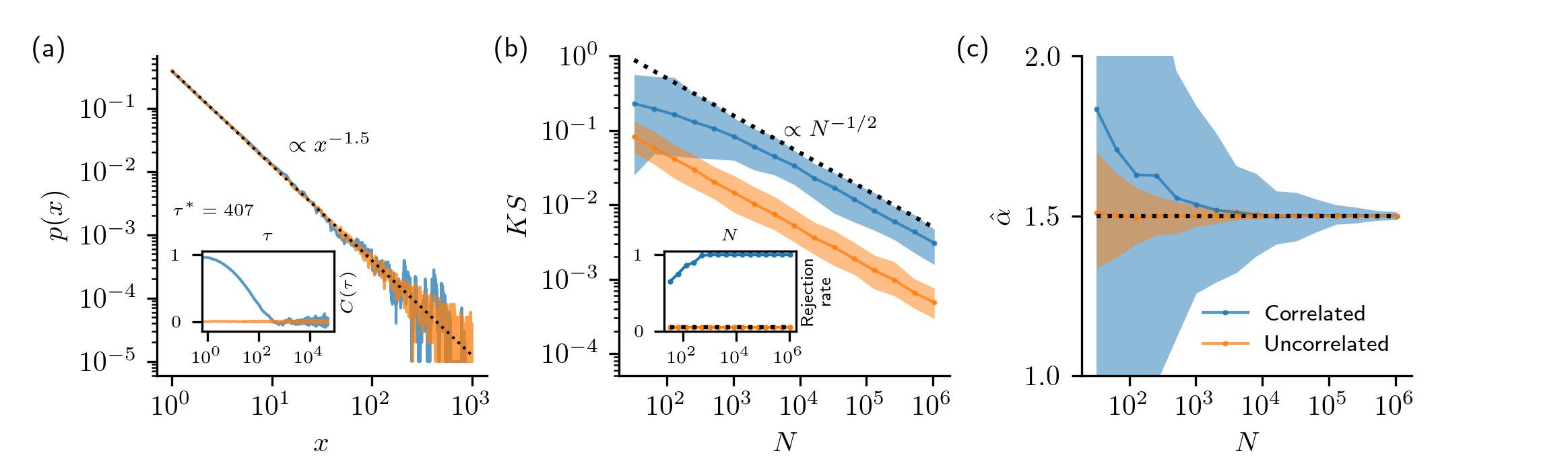}
\caption{
Correlations impact the fitting of power-law distributions using Maximum Likelihood methods.
Two synthetic datasets following a power law with exponent $\alpha=1.5$ for $x=1,\ldots,1000$ were generated: one using independent sampling (in orange /light~gray) and one with correlations (in blue/dark~gray); see SM, Sec.~II for different choices of the maximum cutoff leading to similar results. 
(a) Distribution $p(x)$ for a single realization with $N=10^5$. Inset: Autocorrelation function $C(\tau)$. 
(b) Average and $95\%$ confidence interval of the KS-distance over $100$ different realizations of the synthetic data. Inset: Rejection rate, i.e. fraction of realizations for which the power law is rejected on a 0.05-significance level according to method of Ref.~\cite{CSN} (dotted line) for datasets of varying length $N$. 
(c) Average and $95\%$ confidence interval of the estimated powerlaw exponent $\hat{\alpha}$ over $100$ different realizations of the synthetic data.
}
\label{fig.fit.mcmc}
\end{figure*}

In Fig.~\ref{fig.fit.mcmc} we show that the violations of H2 have a strong influence on the analysis of statistical laws formulated in H1. In particular, the application of the traditional recipes~\cite{CSN} lead to the wrong conclusion that the data is not compatible with a power law distribution: the probability of rejecting the null hypothesis at a $5\%$ significance level is much larger than $5\%$ even for small sample sizes $N$ (inset of Fig.~\ref{fig.fit.mcmc}b). This corresponds to a type-I error because, by construction, the data satisfies H1.
The origin of this failure thus originates from the fact that correlations lead to an effective reduction of the number of independent observations implying larger fluctuations which lead to larger deviations from the fitted model. Specifically, 
we recall that the test employed in Ref.~\cite{CSN} consists of comparing the Kolmogorov-Smirnov (KS) distance between the correlated data and the fitted curve, $KS_{correlated}$ (blue curve), and the KS distance  between independent samples of the model (H1+H2) and the fitted curve, $KS_{model}$ (orange curve). More precisely, the statistical law is rejected at $5\%$ significance level if $KS_{correlated}  > KS_{model}$ in $95\%$ realizations (samplings) of the model.  While in our artificial data  $KS_{correlated} \propto 1/\sqrt{N}$ (as expected) and thus $KS_{correlated} \rightarrow 0$ for $N \rightarrow \infty$, this convergence is shifted from the convergence of $KS_{model}$ (Fig.~\ref{fig.fit.mcmc}b) due to the correlations. This shift leads to an increased rejection rate ($\approx 1$, p-value $\approx 0$).

Violations of H2 are important not only in the hypothesis-testing setting discussed above, they also lead to increased systematic and statistical errors (bias and fluctuations) in the fitting of the parameter $\hat{\alpha}$ (Fig.~\ref{fig.fit.mcmc}c) and, thus, in the selection between models~\cite{Hastie09,Burnham02}.

\paragraph*{Real data.}  In order to confirm that the results discussed above are also relevant in real datasets -- which have a fixed size $N$ -- we consider two types of undersampling of data to sizes $n<N$: taking $n$ points either randomly or preserving the structures/correlations by taking consecutive portions of the time series  (the network and word-frequency databases are first mapped to a time series, as in Fig.~\ref{fig.autocorr}). 
In order to distinguish between the effect of the shape of the distribution (H1) and correlations (H2) we compare the distribution of the $n$ points with i) the proposed statistical law and ii) the empirical distribution (i.e., the one obtained for $n=N$). 
Our results (see SM, Sec.~IV) confirm that correlated data show higher rejection-rate and fluctuations of parameters.

\begin{figure*}[bt]
\includegraphics[width=2\columnwidth]{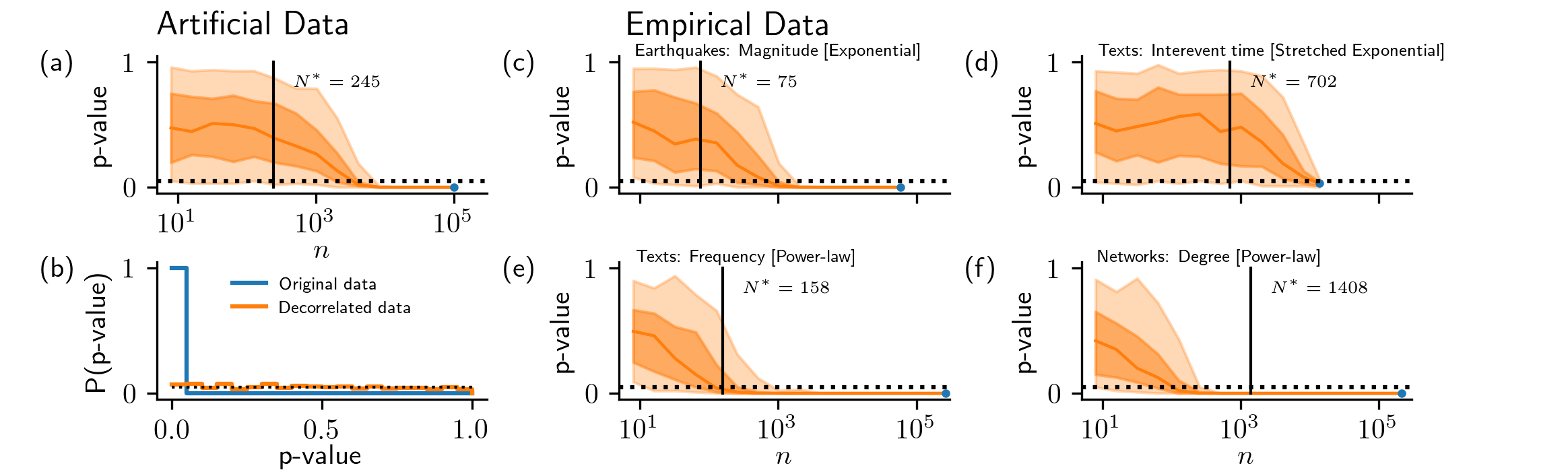}
\caption{
  Decorrelating data leads to different conclusions in hypothesis testing in artificial (a,b) and empirical (c-f) data. The distribution of p-values from fitting correlated (of size $N$, in blue) and subsampled  shuffled data (of size $n \le N$, in orange). While the correlated data leads to a peaked distribution of small p-values (i.e. rejection), the decorrelated data obtained from our approach leads to the expected flat distribution of p-values. The effective sample size $N^*$  (black vertical line) was obtained from $\tau^*$ reported in Fig.~\ref{fig.autocorr} as $N^*=N/\tau^*$. While all cases are rejected when fitting the full dataset, in 3 out of 4 cases we cannot reject the null hypothesis for decorrelated data (median p-value $\geq 0.05$ at $n=N^*$).}
\label{fig.fit.real}
\end{figure*}

\paragraph*{Alternative approach.}
In the vast literature of statistical methods for dependent data, two general approaches can be identified. The first approach is to incorporate the violation of independence in more sophisticated (parametric) models, e.g., in time series one could consider Gaussian/Markov processes~\cite{vankampen}. This is of limited use in our case because statistical laws aim to provide a coarse-grained description (stylized facts) valid in many systems, instead of different detailed models of particular cases. The second (non-parametric) approach, which we pursue here, is to de-correlate or de-cluster the data, leading to a dataset with an ``effective'' sample size $N^*\le N$~\cite{Gasser,Weiss78,Chicheportiche11}. In practice, the analysis consists of multiple realizations of the following three steps: 
 \begin{itemize}
 \item[(i)] Randomize (shuffle) the original sequence and select randomly $n$ points, for different $n \in [1,N]$.
 \item[(ii)] Apply the traditional statistical analysis  (i.e., the hypothesis test, model comparison, and fitting based on H1+H2) to the randomized dataset obtained in (i), investigating their dependence on $n$.
 \item[(iii)] Estimate the correlation $\tau^*$, defined as the time after which two observations (in the time series) are independent from each other. Out of the total $N$ samples we thus estimate $N^*=N/\tau^*$ to be the number of independent samples and therefore we select the results from step (ii) for $n \approx N^*$. 
 \end{itemize}

 The determination of $\tau^*$ -- or the effective sample size $N^*$ -- in step (iii) requires knowledge or assumptions about how the data was generated. For the case of temporal sequences we propose to compute the auto-correlation and take as $\tau^*$ the lag for which it reaches an interval around zero (1-percentile of the random realizations, as in Fig.~\ref{fig.autocorr}). In the constructed example (Fig. 2), we obtain $\tau^*=407 \Rightarrow N^* = 245$ which leads to a rejection-rate (at p-value$=0.05$) equal to $5\%$ for all $n<N^*$. For the case of networks, the determination of the effective sample size $N^*$ depends on the generative process and/or the sampling used to measure the data (here we assumed a specific edge sampling method, as described in Fig.~\ref{fig.autocorr}.)
 In Fig.~\ref{fig.fit.real} we show evidence of the effectiveness of our approach through a systematic analysis of the p-value distribution as a function of $n$ for both the constructed and empirical datasets. This is further corroborated in artificial data (see SM, Sec.~V) showing  that i) our method for the selection of $\tau^*$ is superior to the one proposed in Ref.~\cite{Gasser} (sum of the autocorrelation function) and ii) can be equally applied to data with other types of correlation: a Markov process with negative correlation and a Gaussian process with long-range correlations. In all cases our approach shows an uniform distribution of p-values under the null hypothesis.

An important message of our analysis is that conclusions about the statistical law can be obtained even when the precise value of $\tau^*$ (or the effective sample size $N^*$) is unknown in step (iii).  By shuffling and undersampling the sequence at different sizes $n$ -- steps (i) and (ii) -- we can investigate how the results depend on $n$ and obtain the range in $\tau^*$ for which the different conclusions hold.
  For instance, in the case of earthquakes -- Fig.~\ref{fig.fit.real}c -- we see that the rejection increases dramatically around $n \approx 10^3$. We thus conclude that, in this dataset of size $N \approx 10^5$, we falsify the Gutenberg-Richter law if  $\tau^* \le N /n \approx 10^2 \text{ observations } \approx 20$ days~\cite{Note3}. The conservative estimate of $\tau^*$ in Fig.~\ref{fig.autocorr} was $\tau^*=791 > 10^2$ and therefore we conclude that based on this data we cannot reject the Gutenberg-Richter law, contrary to the conclusion obtained assuming independent observations.  We find similar results -- Fig.~\ref{fig.fit.real}d -- for the stretched exponential distribution of inter-event times between words, while for Zipf's law -- Fig.~\ref{fig.fit.real}c -- the outcome is uncertain, and the power-law degree distributions in networks -- Fig.~\ref{fig.fit.real}d -- is rejected even in the correlated case.

\paragraph*{Discussion and Conclusion}

Statistical laws in complex systems are typically formulated (as in H1) without reference to the generative process of the data.
Therefore, ideally, the empirical test of these laws should be designed to account for a large class of processes generating $\{x_i\}$. Traditional methods~\cite{CSN} based on the hypothesis of independent data (H2) are weak tests because they include a strong hypothesis that is easily violated, therefore favoring rejection. In fact, here we have shown how these methods: (i) lead to wrong rejections of the laws because of correlated data; and (ii) are over-optimistic regarding uncertainties of the estimated parameters. Stronger tests of statistical laws should make weaker assumptions about the generative process so that rejections of the compound hypothesis provide much stronger evidence of the rejection of the law (H1). Here we proposed a methodology which allows us to identify the strongest assumption about correlations of the data $\tau^*$ for which the law can be rejected.
Being conservative in the choice of $\tau^*$ (i.e., choosing large values for which we are confident that $x_i$ and $x_{i+\tau^*}$ are uncorrelated)
overcomes the main shortcoming of the traditional approach~\cite{CSN} and
ensures that when we reject the law this is not happening due to correlations in the data (failing to reject the law is never a confirmation of its validity). 
In this sense, our approach is similar  in spirit to the Bonferroni correction to account for multiple hypothesis testing~\cite{bonferroni} (both aim to avoid overconfident or spurious rejections of hypotheses).
Our approach is even applicable in cases with no well-defined mixing time~$\tau^*$ (e.g. long-range correlations) because it yields very large values of $\tau^* \approx N$ (no two points are independent).

Instead of directly testing whether the statistical law is valid (hypothesis testing), often the best we can do is to compare different alternatives (model comparison)~\cite{StumpfEPL,Broido,CSN,Laws,PRX,Hastie09,Burnham02}. Also in this case violations of the hypothesis of independence are important and have been mostly ignored in the analysis of statistical laws in complex systems (see Refs.~\cite{Lee,Khanin,Sornette} for exceptions). As shown above, due to correlations (and violations of H2) actual data show much larger fluctuations than expected under the hypothesis of independent observations. By using a shuffled and undersampled dataset we obtain larger uncertainties in the estimated parameters; we expect similar lack of certainty in the choice of best models. The need to account for violations of the independence assumption, shown in this Letter, applies much more broadly than the cases treated above. Correlations should be accounted for whenever testing statistical laws in complex systems, such as linguistic laws~\cite{Laws}, scaling laws with system size -- maximum likelihood methods based on H2 have been applied to biological allometric laws~\cite{dodds} and to city data~\cite{us} -- and different distributions of inter-event time (burstiness)~\cite{Bunde,Barabasi,bursts,burstsC}. 

\paragraph*{Acknowledgments:}
We thank F. Font-Clos and J. Moore for the careful reading of the manuscript. 

\paragraph*{Code and data availability:}see Ref.~\cite{codes}.



\newpage

\setcounter{equation}{0}

\setcounter{figure}{0}
\renewcommand{\figurename}{SM-Fig.}

\begin{widetext}
  \begin{center}
{\Large Supplementary Material }
\end{center}

\section{Fitting}

Given a set of data $\left \{ x_i \right \} = x_1,x_2,\ldots, x_N$ we find the best parameters $\hat{\vec{\alpha}}$ by maximizing the log-likelihood:
\begin{equation}
 \log \mathcal{L} = \sum_{i=1}^N \log p(x_i;\vec{\alpha}).
\end{equation}

\paragraph*{Exponential.} 
For the earthquake data, the observable $x$ is the magnitude. 
We thus fit a continuous exponential distribution with support $x \in [x_{\min},\infty)$
\begin{equation}
 p(x; \vec{\alpha}) = p(x,\alpha) = \alpha e^{-\alpha (x-x_{\min})},
\end{equation}
where $x_{\min}$ is a constant.
We choose $x_{\min}=2$ according to Ref.~[29] main text.

\paragraph*{Stretched Exponential.}
For the interevent times data, the observable $x$ is the number of words between two consecutive occurrences of a given word (e.g. ``the'') in a book.
We fit a discrete stretched exponential with two parameters $\vec{\alpha} = (\alpha,\beta)$ and support $x=x_{\min},x_{\min}+1,\ldots,\infty$ defined by its cumulative distribution
\begin{equation}
 F(x; \vec{\alpha}) \equiv P(X \geq x;\vec{\alpha}) = e^{-\alpha (x^{\beta} - x_{\min}^{\beta}) }.
\end{equation}
From this we obtain the probability distribution as $P(x;\vec{\alpha}) = F(x;\vec{\alpha}) - F(x+1;\vec{\alpha}) $.
We choose $x_{\min}=3$ as the minimum value for which estimation of the parameters is approximately independent of $x_{\min}$ (not shown).

\paragraph*{Power law.}
For the text and network data, the observable $x$ is a rank. 
We thus fit a discrete power-law with support $x =1,2,\ldots, V $ with $V$ the maximum (observed) rank and $\alpha>0$ such that
\begin{equation}
 p(x; \alpha) = C x^{-\alpha}
\end{equation}
with $C = \frac{1}{H_{V,\alpha}}$, where $H_{V,\alpha} \equiv \sum_{x=1}^V x^{-\alpha} $ is the generalized harmonic number of order $\alpha$ of $V$.

\section{Edge-sampling networks}

Here we describe the procedure used to obtain an ordered sequence of nodes $x_i, i=1, \ldots, 2 L$ starting from a simple graph of $V$ nodes and $L=N/2$ edges (or links).  We select nodes by choosing the edges $x_i \leftrightarrow x_j$ of a network in a particular order (without replacement) and then adding the two nodes linked by this edge ($x_i,x_j$)  in a random order to the list of observations $x_i$.  A node with degree $k$ thus appears $k$ times in the sequence. If the node labels correspond to their ranks (in degree), $p(x)$ corresponds to the rank frequency distribution. 

We start from a randomly selected edge. The next edge is selected randomly from the remaining list of edges involving $x_i$ or $x_j$ (with probability $0.5$) or randomly from the complete list of remaining edges in the network (with probability $0.5$). If there are no remaining edges involving $x_i$ or $x_j$ we choose an edge randomly in the complete network. This corresponds to a combination of random sampling of edges with a random walk (using local steps) in the network. This procedure is repeated until all edges $L$ of the network are sampled, leading thus to a sequence $x_i$ for $i=1, \ldots, 2L$.

Lee et al.~[SM4] showed that random edge sampling is equivalent to node sampling (previously discussed in Ref.~[SM5]), both leading to deviations of the degree distribution from power-law. The local steps used here enhance the extent into which the measured $x_i$ deviate from an independent sample (H2), as shown in Fig.~1d of the manuscript. Correlations in the sequence $x_i$ exist even forbidding local steps because when we sample edges we sample two nodes and the selection of the two is not random, e.g., degree-degree correlation in networks. Another distinction of our approach is that we work in the rank-frequency picture  so that node $x$ with degree $k_x$ contributes with $k_x$ symbols $x$ in the sequence $x_i$ (in contrast to the analysis of the distribution $P(k)$ of nodes of degree~$k$).

\section{Artificial Data}\label{sec.artificial}

Here we describe how to generate a correlated sequence $x_1, x_2, \ldots, x_i, \ldots, x_N$ with variable correlation time $\tau^*$ and an arbitrary marginal distribution $p(x)$ for $N \rightarrow \infty$.
We construct a Markov process of order one $p(x_{i+1}|x_i, x_{i-1},\ldots) = p(x_{i+1}|x_i)$ such that the next value $x_{i+1}$ is proposed from $p(x)$ with probability $\mu$ and from a pre-defined set of neighbours $B(x_i)$ with probability $1-\mu$:
\begin{equation}\label{eq.prop}
  p(x_{i+1}|x_i) = \begin{cases}
    p(x; \alpha)  & \text{ w/ probab. } \mu\\
      \frac{1}{|B(x_i)|} \delta(x_i \in  B(x_i))  & \text{ w/ prob. } 1-\mu,
    \end{cases}
\end{equation}
where $|B(x_i)|$ is the number of elements of $B(x_i)$.
The strict validity of $p(x)$ is ensured by accepting or rejecting this proposal using the Metropolis-Hasting method~[SM1]. Starting from a random value $x$, a new value $x'$ is proposed according to Eq.~(\ref{eq.prop}). This proposal is accepted  with a probability given by the Metropolis-Hasting condition
$$ A (x\mapsto x') = \frac{g(x' \mapsto x) p(x')}{g(x \mapsto x')p(x)},$$
where $g(x \mapsto x')$ is the probability of proposing $x'$ from $x$, and $p(x)$ is the probability distribution we wish to impose for $x$ (e.g., the power-law distribution). If $A>1$, $x'$ is accepted. If $x'$ is accepted, $x_{i+1}=x'$, otherwise $x_{i+1}=x_i$. Each $x$ has neighbours $B(x)$  that we assume to be reciprocal, i.e. $x' \in B(x) \Rightarrow x \in B(x')$ for all $x,x'$. This implies that $A=1$ if $x$ and $x'$ are not neighbours of each other. If they are neighbours, the probabilities $g$ obtained from the rule~(\ref{eq.prop}) are
$$  g(x\mapsto x') =\mu p(x')+(1-\mu)/|B(x)|.$$
The procedure above is repeated $N$ times. The autocorrelation of $x$ decays exponentially and $\frac{\# x}{N} \rightarrow p(x)$ for $N\rightarrow \infty$. For $\mu=1$ the data becomes independently distributed in agreement with H2. For $\mu < 1$ the sequence of $x_i$ is correlated with a correlation time $\tau \lessapprox 1/\mu$. In the example shown in the manuscript we consider: $p(x)$ to be a power-law distribution -- Eq.~(1) of the paper -- with support $x=1,\ldots,1000$ (with qualitatively same results if other maximum cutoffs are chosen, see Fig.~\ref{fig.synth.cutoff}), $\mu =0.01$, and the neighbours $B(x_i)$ of $x_i$ in the local step $x_{i+1}$ to be the $k-$nearest neighbours $x_i-k, \ldots, x_i-1, x_i, x_i+1, \ldots x_i+k$ (i.e., we choose one of the $k$-nearest neighbours by chance with equal probability) setting $k=5$. 

The process described above is a generic process that aims to capture some of the features of different systems.
In a text, the neighbours of $x_i$ could correspond to the word types which are syntactically allowed to follow $x_i$. In a network for which nodes (and their degrees $x_i$) are discovered through a random walk through its links, the step of choosing neighbours of $x_{i}$ mimics the cases in which nodes connected to a given node have a degree more similar to it than a random node (positive degree assortativity of the underlying network) leading to $p(x_{i+1}\approx x_i |x_i) > p(x_i;\alpha)$.

\section{Separating correlations and models in fitting real data}

Here we show how correlations in real data affect the fit of different models similar to the analysis on synthetic data.
In contrast to synthetic data, for real data we do not know the generating process and thus we don't know the underlying distribution and we are unable to generate datasets of arbitrary length $n<N$.  
Therefore, when fitting a model to the full data, it is impossible to know whether rejection of the hypothesis is due to correlations or due to deviations from the underlying distribution.
In order to disentangle the two effects, isolating the effect of correlations, we consider two approaches. First, we shuffle and undersample the data randomly (as argued in the main manuscript). Second, we compare the data not only to the parameterized statistical law but also to a 0-parameter function defined by the empirical distribution of the full data set.

The results for the four datasets used in the main manuscript are shown in SM-Figs. \ref{fig.fit.real.earthquakes}, \ref{fig.fit.real.intereventtimes}, \ref{fig.fit.real.texts}, and \ref{fig.fit.real.networks}.
We calculate the KS distance between the empirical distribution and correlated and uncorrelated datasets, respectively, of different size obtained from subsampling (panel a).
We observe that the KS-distance for the correlated data is much larger than for the uncorrelated data. 
While the uncorrelated data is by construction rejected at a rate of $0.05$, the correlated data is rejected with a much higher probability. This confirms that violations of the hypothesis of independent data (H2) lead to rejections. Fitting the parametric statistical law leads to qualitatively similar results (panel b). The correlated data yields much larger values of KS-distance to the fitted distribution than the uncorrelated data. This in turn leads to an increase in the rejection rate of the model. Beyond hypothesis testing, we also observe that the correlations have a substantial impact on the estimation of the parameters of the respective model (panel c) -- not only in terms of an increase in the fluctuations but also in the average value.

\section{Testing our method in artificial data}

Here we report numerical tests of the method we propose in the main manuscript applied to artificially-generated data. We compare the outcome of our method to the method proposed by Gasser~[SM2]

The artificial data (described in Sec.~III above) corresponds to a Markov process with stationary distribution $p(x;\alpha)$. Starting from an arbitrary initial condition $x_i$, the probability that after $t$ times the state $x_{i+t}$  equals $x$ approaches the stationary distribution $p(x;\alpha)$ for long times $t$, with the typical time for this to occur given by the mixing time~$\tau_m$ of the Markov Chain~[SM3]. Points separated by times $t>\tau_m$ can thus be considered as independent samples from $p(x;\alpha)$. The method we propose selects randomly $n$ out of the $N$ points in the data, implying that these points are separated by a distance $N/n$. In our method, the specific value of $n$ chose as the effective sample size $n=N^*$ is determined by the correlation time $\tau^*$ -- computed from the autocorrelation function of $\{x_t\}$ -- so that $N/N^* = \tau^*$. Our goal of having an effective sample size of independent points is achieved if  $\tau^* > \tau_m$, while ideally $\tau^* \approx \tau_m$.

The reasoning above indicates that the identification of the time $\tau^*$ is a critical element of our method. The alternative method we consider here, by Gasser~[SM2], proposes an alternative estimation of $\tau^*$ as
\begin{equation}\label{eq.gasser}
  \tilde{\tau}^* = \sum_{\tau = -\infty}^{\infty} (C(\tau))^2,
  \end{equation}
where $C(\tau)$ is the auto-correlation function of the series $x_i$ at lag $\tau$.

We first consider the artificial data described in the previous section for different values of the free-parameter $\mu$. SM-Figure~\ref{fig.fit.synth.test} shows the comparison between the traditional approach (ignoring correlations),  our method, and Gasser's method. 
The superiority of our method is confirmed by the fact that we obtain a flat distribution of p-values (panel a-c), expected from the fact that by construction the stationary distribution of the Markov Chain equals the fitting distribution.
In order to quantify the deviation from a flat distribution, we calculate the entropy over the distribution of p-values varying the correlation strengths $\mu$ (panel d).
Our method consistently yields an entropy virtually indistinguishable from the maximum value expected from a flat distribution (dotted line).
In contrast, the other methods yield substantially smaller entropy values, reflecting a deviation from a flat distribution.
This in turn translate into much smaller or larger fraction of realizations being rejected than expected from the imposed significance threshold, i.e. p-value $<0.05$ (panel e).
The main reason for the best performance of our method is that the estimated correlation time $\tau^*$ is larger than the one estimated by the alternative method in Eq.~(\ref{eq.gasser}) (see panel f).

Our method applies also to datasets showing different types of auto-correlation functions. To confirm this, we performed comparisons -- similar to the ones reported above -- using other types of artificial data that have a known marginal distribution $p(x;\alpha)$ but different types of autocorrelation. 
For instance, in SM-Fig. \ref{fig.synth.anticorr} we test power-law data with negative correlations (for odd time lags) and in  SM-Fig.~\ref{fig.synth.lrc} we test Gaussian data with long-range correlations.
Even though both samples follow by construction the tested statistical law, the distribution of p-values is peaked at very small values, leading to a much larger rate of rejection than expected by chance. 
Applying our methodology (i.e., estimating a correlation time $\tau^*$ from the auto-correlation, shuffling, and subsampling the data to a smaller size $n=N^*=N/\tau^*$) yields an approximately uniform distribution of p-values consistent with the validity of the null hypothesis.

\section*{References}

\begin{itemize}
\item[[SM1]] C.P. Robert, G. Casella, Monte Carlo statistical methods,
  Springer texts in statistics, 2nd ed. (Springer, Berlin, 2005), ISBN 0387212396

\item[[SM2]] T. Gasser, {Goodness-of-{Fit} {Tests} for {Correlated} {Data}}, Biometrika {\bf 62}, 563 (1975).
  
\item[[SM3]] D. A. Levin and Y. Peres, {\it Markov Chains and Mixing Times}, American Mathematical Society (2017).

\item[[SM4]] S. H. Lee, P.-J. Kim, and H. Jeong, Statistical properties
  of sampled networks, Phys. Rev. E 73, 016102 (2006).
  
\item[[SM5]] M. P. H. Stumpf and C. Wiuf, Sampling properties of
random graphs: The degree distribution, Phys. Rev. E
72, 036118 (2005).

\end{itemize}

\newpage

\begin{figure}[bt]
\includegraphics[width=1\columnwidth]{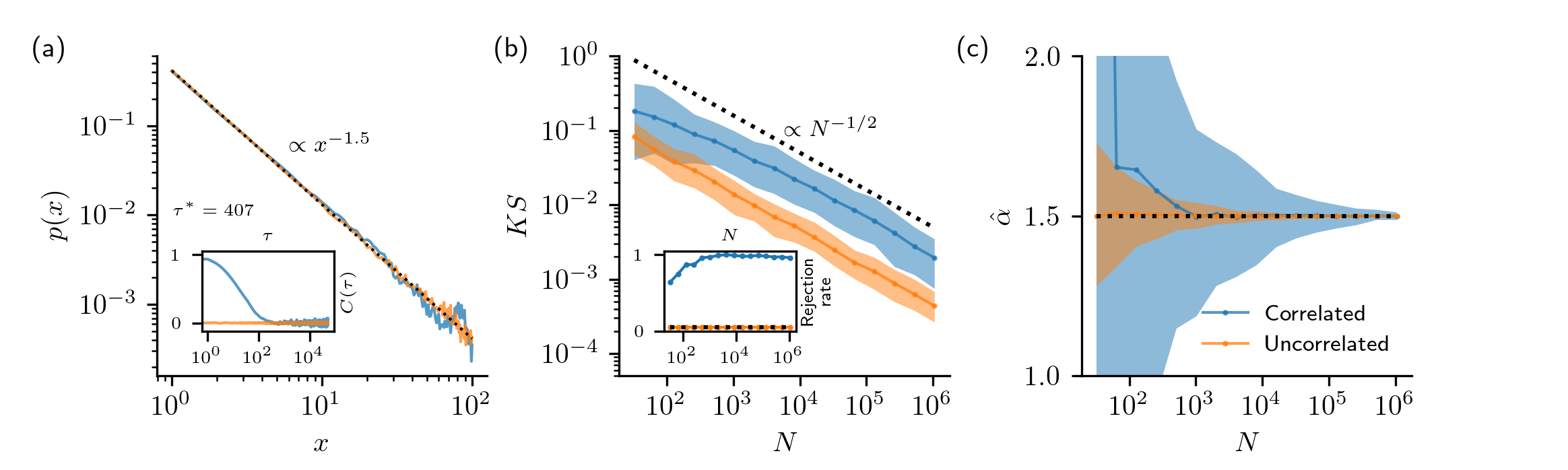}\\
\includegraphics[width=1\columnwidth]{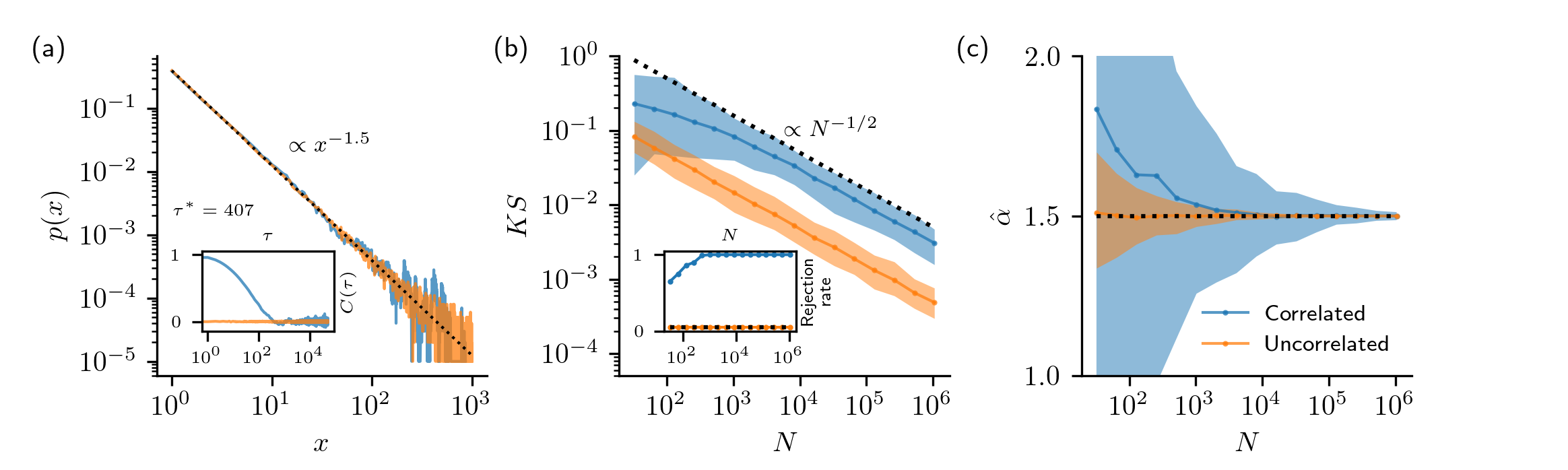}\\
\includegraphics[width=1\columnwidth]{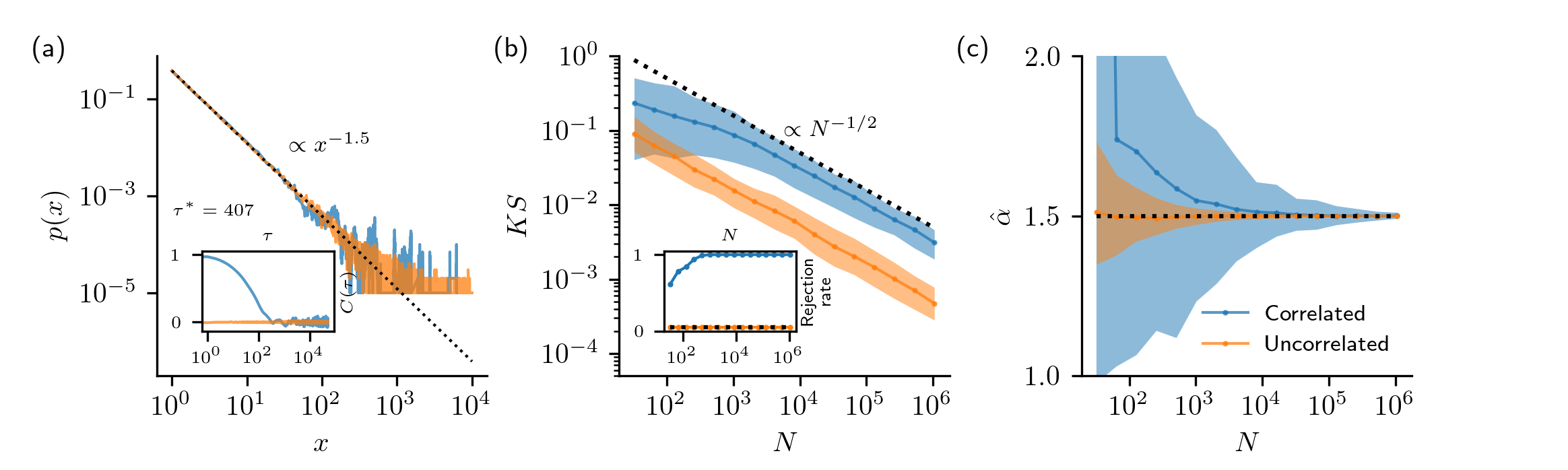}
\caption{
Results from Figure 2 using synthetic data are robust with change of maximum cutoff in the power-law: $V=10^2$ (top row), $V=10^3$ (middle row), and $10^4$ (bottom row).
In each row we show two synthetic datasets following a power law with exponent $\alpha=1.5$ for $x=1,\ldots,V$: one using independent sampling (in orange /light~gray) and one with correlations (in blue/dark~gray).
(a) Distribution $p(x)$ for a single realization with $N=10^5$. Inset: Autocorrelation function $C(\tau)$. 
(b) Average and $95\%$ confidence interval of the KS-distance over $100$ different realizations of the synthetic data. Inset: Rejection rate, i.e. fraction of realizations for which the power law is rejected on a 0.05-significance level according to method of CSN (dotted line) for datasets of varying length $N$. 
(c) Average and $95\%$ confidence interval of the estimated powerlaw exponent $\hat{\alpha}$ over $100$ different realizations of the synthetic data.
}
\label{fig.synth.cutoff}
\end{figure}

\begin{figure}[bt]
\includegraphics[width=1\columnwidth]{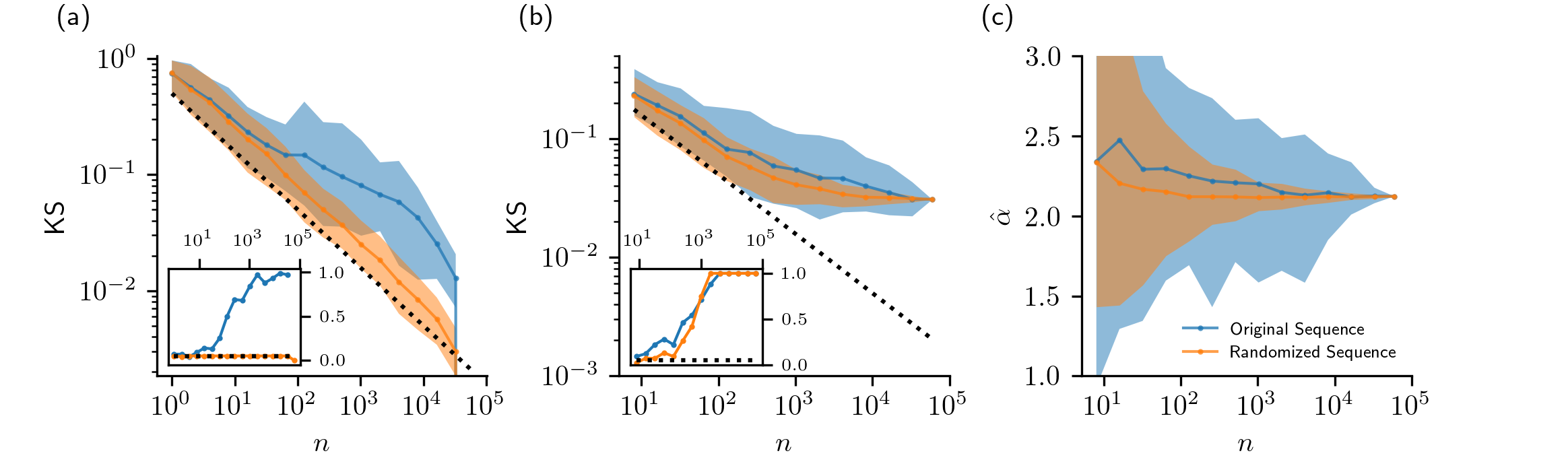}
\caption{
    Effect of correlations on fitting for earthquake data.
  \textit{Original Sequence} selects an undersample of $n$ consecutive observations from the original sequence of size $N$ (assuming periodic boundary conditions);  \textit{Randomized Sequence} selects a random undersample of $n$ non-consecutive observations from the original sequence (without replacement). 
  (a) KS-distance between the undersampled data and the empirical distribution from full data. Inset: Rejection rate, fraction of realizations for which the empirical distribution is rejected on a 0.05-significance level (dotted line) for undersampled datasets of varying length $n$. 
  (b) Inset: KS-distance between undersampled data and fitted exponential distribution from Maximum Likelihood estimation.  Inset: Rejection rate, fraction of realizations for which the power-law distribution is rejected on a  0.05-significance level (dotted line) for undersampled datasets of varying length $n$.  
  (c) Estimated power-law exponent $\hat{\alpha}$. Dotted line in (a) and (b) indicate $\propto n^{-1/2}$.
}
\label{fig.fit.real.earthquakes}
\end{figure}

\begin{figure}[bt]
\includegraphics[width=1\columnwidth]{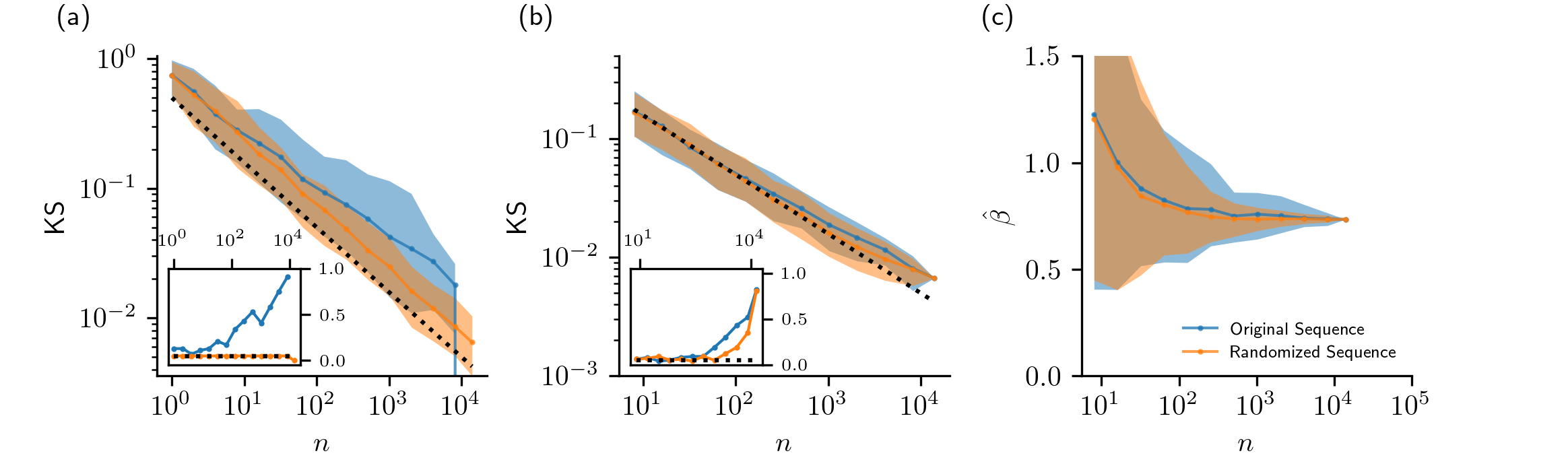}
\caption{
    Same as in  SM-Fig.~\ref{fig.fit.real.earthquakes} for the interevent times data. 
}
\label{fig.fit.real.intereventtimes}
\end{figure}

\begin{figure}[bt]
\includegraphics[width=1\columnwidth]{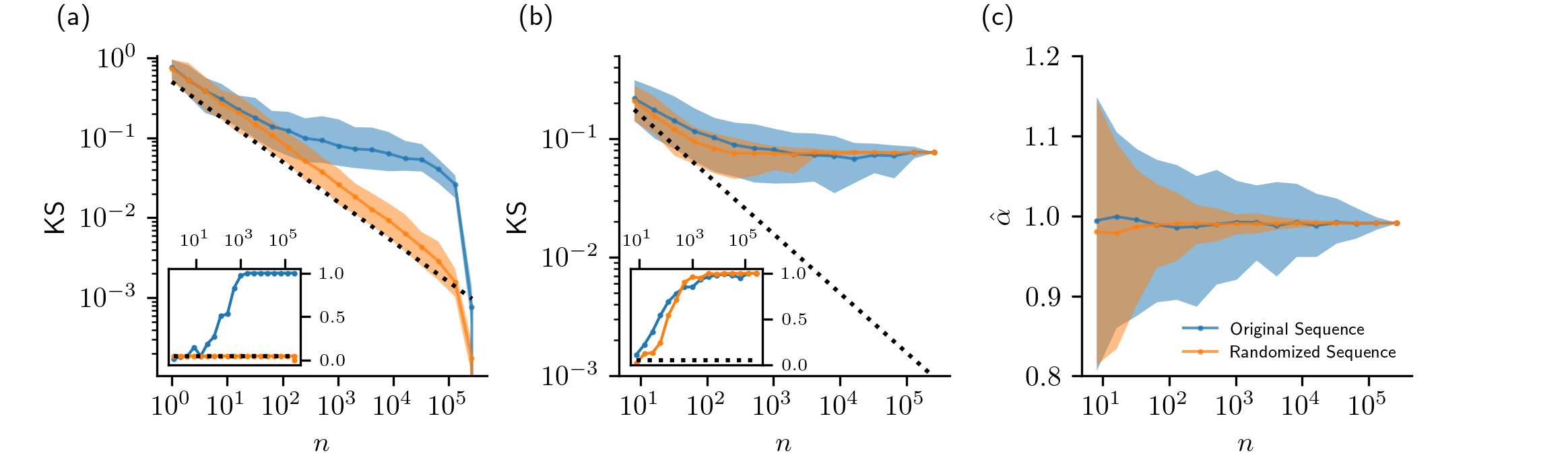}
\caption{
    Same as in  SM-Fig.~\ref{fig.fit.real.earthquakes} for the text data.
}
\label{fig.fit.real.texts}
\end{figure}

\begin{figure}[bt]
\includegraphics[width=1\columnwidth]{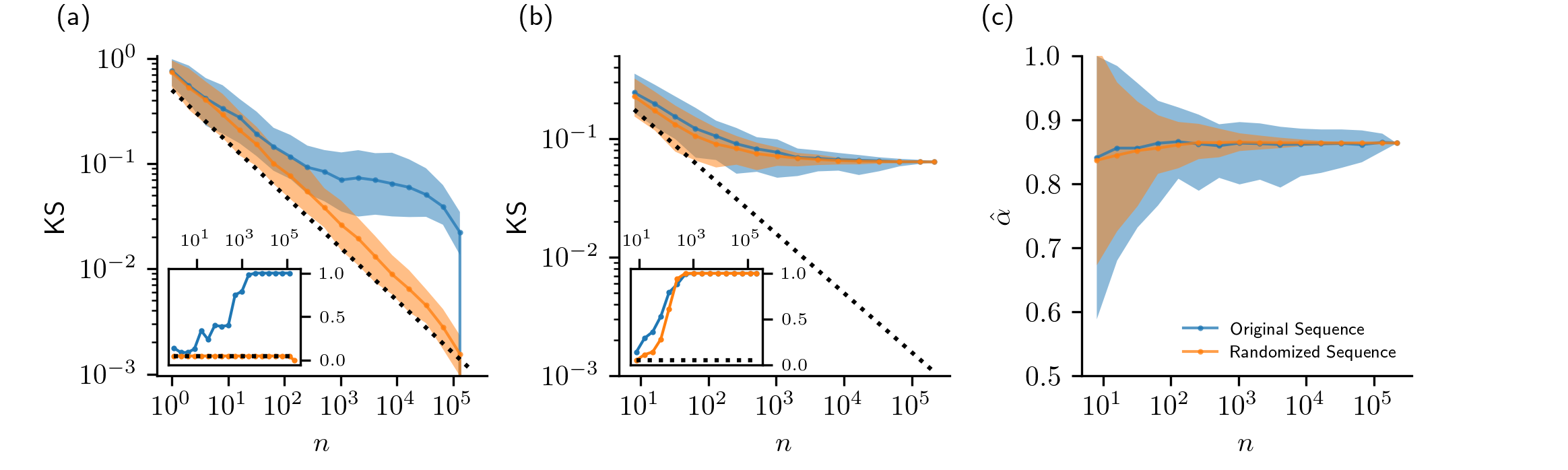}
\caption{
    Same as in  SM-Fig.~\ref{fig.fit.real.earthquakes} for the network data.
}
\label{fig.fit.real.networks}
\end{figure}

\begin{figure}[bt]
\includegraphics[width=1\columnwidth]{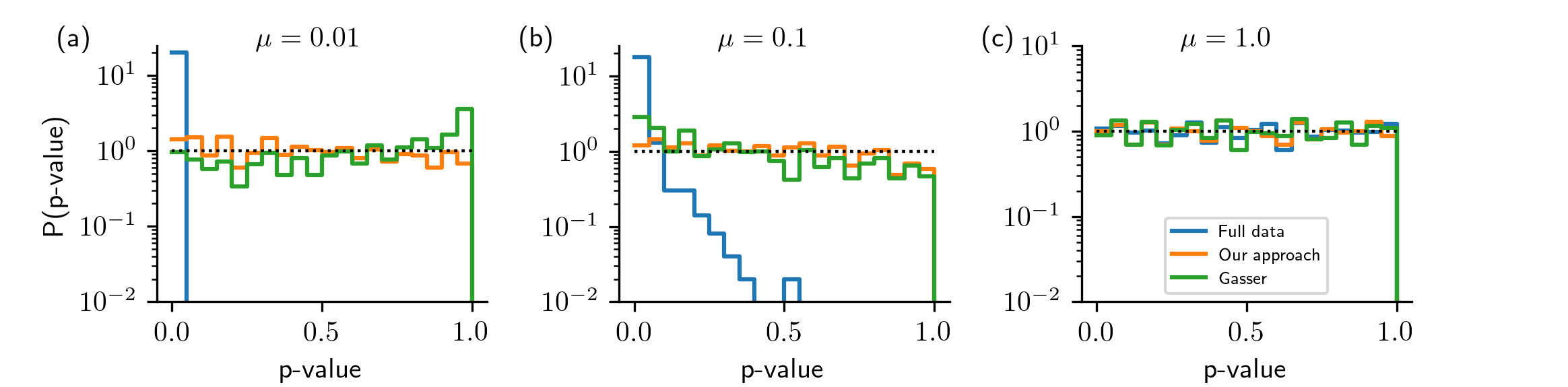}\\
\includegraphics[width=1\columnwidth]{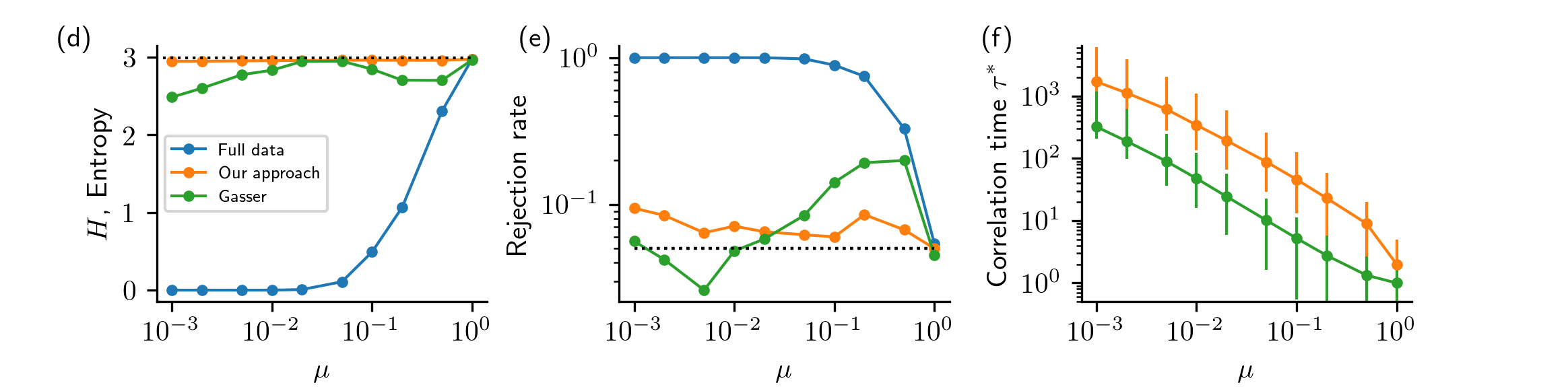}
\caption{
Test of our method to mitigate effects of correlations on hypothesis testing in artificial data. Our method is compared to two other methods (see caption): the traditional approach that ignores correlations (full data) and the one by Gasser [SM2].
(a-c) Distribution of p-values $P(p-\text{value})$ for different values of the parameter $\mu$ used in our artificial data, described in Sec.~III ($\mu$ is proportional to $\tau^*$).
(d) Entropy $H$ of the distribution of p-values as a function of $\mu$. A flat distribution of p-values (ideal outcome) corresponds to $H=3$.
(e) Rejection rate $P(p-\text{value} \leq 0.05  )$ as a function of $\mu$
(f) Estimated correlation time $\tau^*$ as a function of $\mu$.
For the synthetic datasets we use the same parameters as in Figure 2 (main text), i.e. a power law with exponent $\alpha=1.5$ for $x=1,\ldots,1000$ with $N=10^5$ and variable correlation strength $\mu$.
Distributions are obtained from $1,000$ different datasets.
}
\label{fig.fit.synth.test}
\end{figure}

\begin{figure}[bt]
\includegraphics[width=0.32\columnwidth]{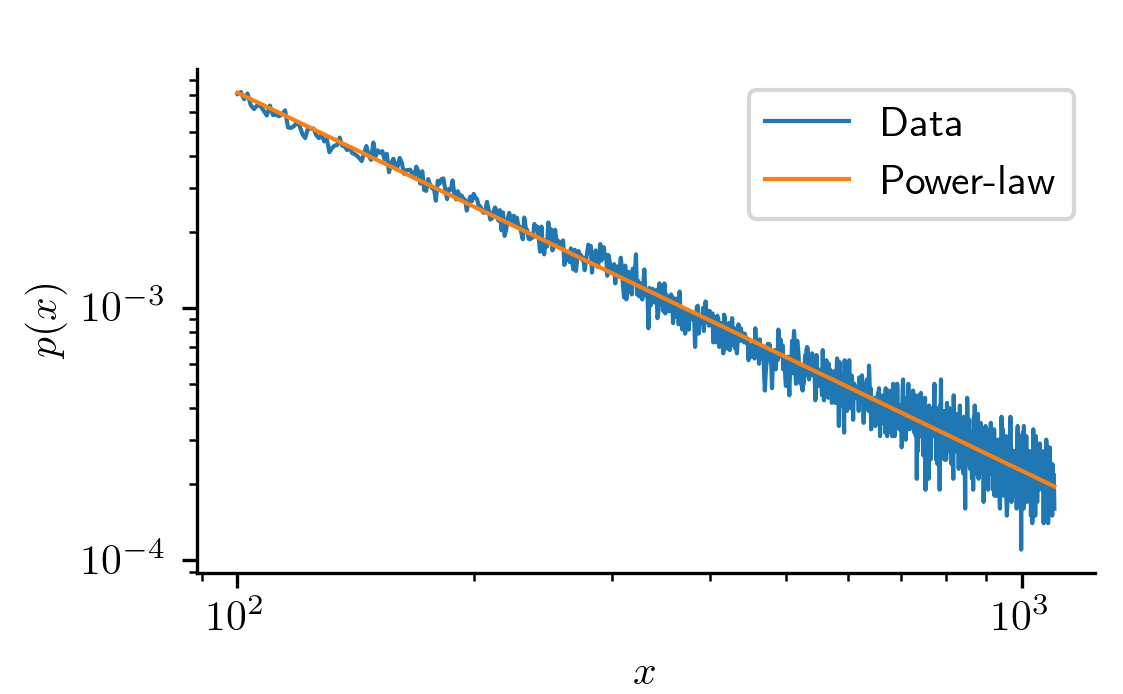}
\includegraphics[width=0.32\columnwidth]{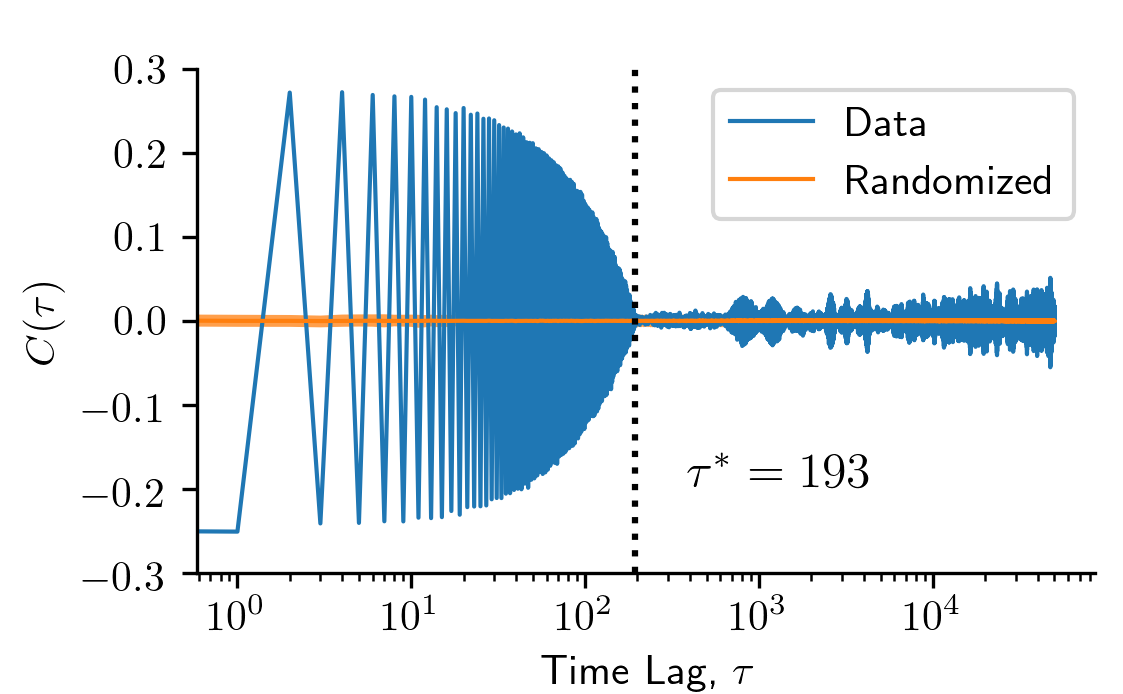}
\includegraphics[width=0.32\columnwidth]{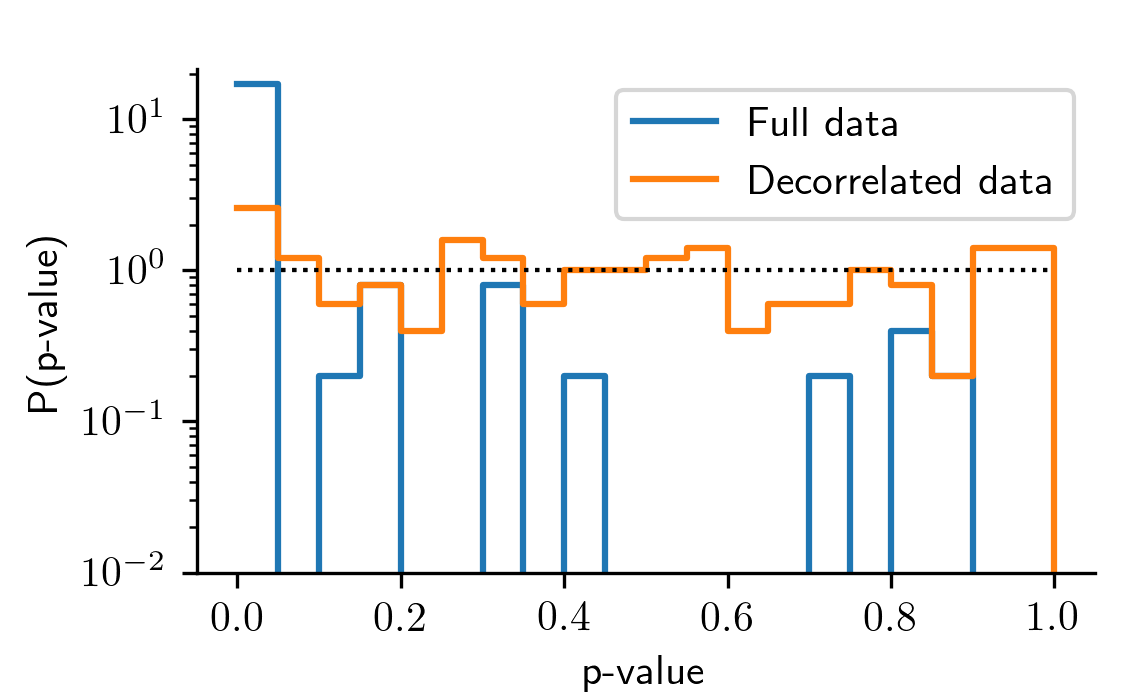}
\caption{
Synthetic data from a power-law distribution with negative correlations (for odd time-lags).
The synthetic data consists of $N=10^5$ samples drawn from a discrete power-law with exponent $\alpha=1.5$ and support $x=100,101,\ldots,1099$.
The correlations are generated by splitting the support into 4 adjacent regions (i,ii,iii,iv) with approximately similar probability. The values of $x$ inside each region are drawn randomly, but the order of the regions is not. We draw either either $200$ samples alternating between region i and iii, sequentially, or we draw $200$ samples that either come from region ii or iv.
Left: Distribution $p(x)$ of the empirical data (blue) and the asymptotic power-law (orange).
Middle: Autocorrelation $C(\tau)$ of the empirical sequence of data (blue) and the shuffled sequence (orange).
Right: Distribution of p-values from 100 different synthetic datasets of size $N=10^5$ for the full correlated data (blue) and the decorrelated data by subsampling to $N^*=N/\tau^*$.
}
\label{fig.synth.anticorr}
\end{figure}

\begin{figure}[bt]
\includegraphics[width=0.32\columnwidth]{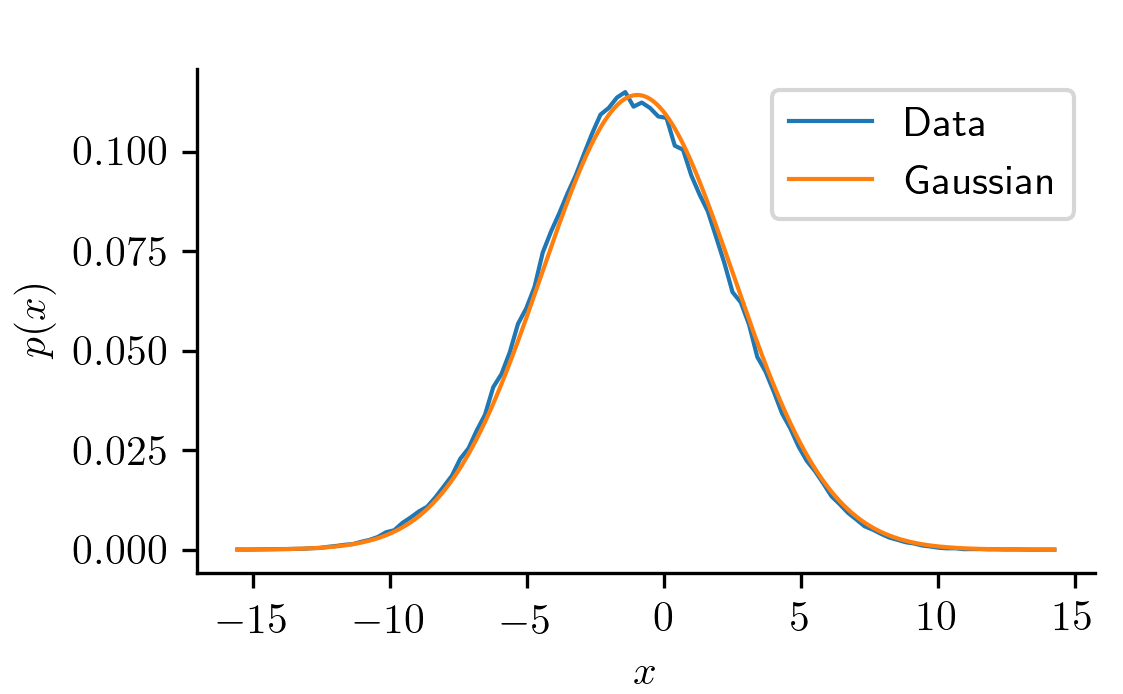}
\includegraphics[width=0.32\columnwidth]{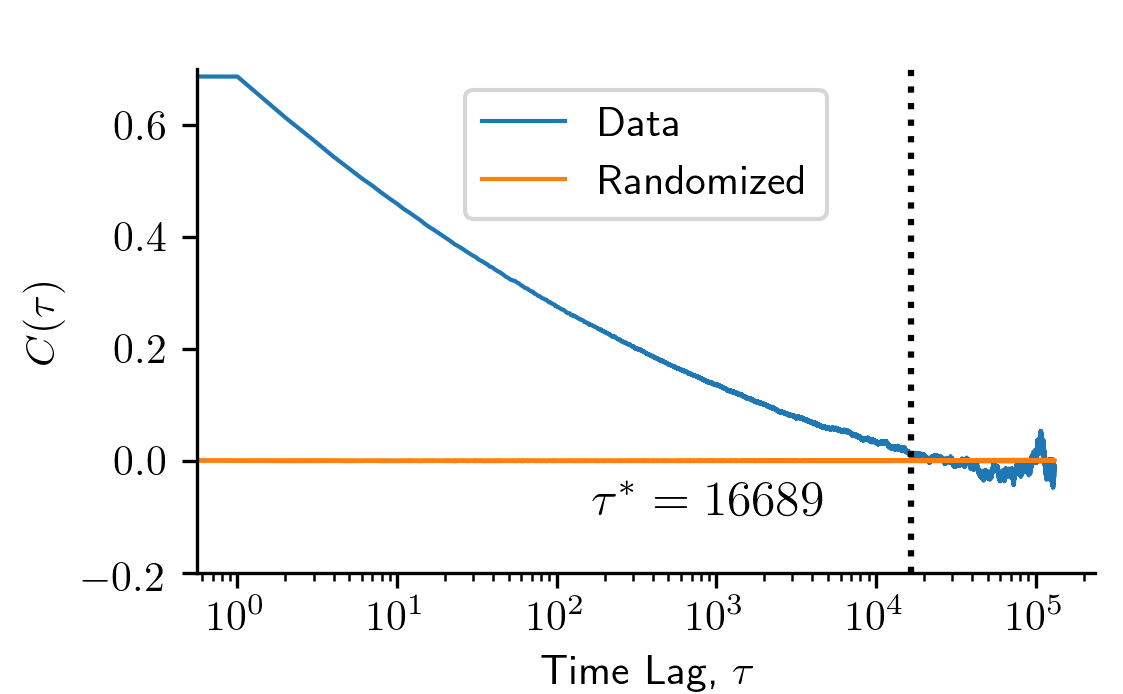}
\includegraphics[width=0.32\columnwidth]{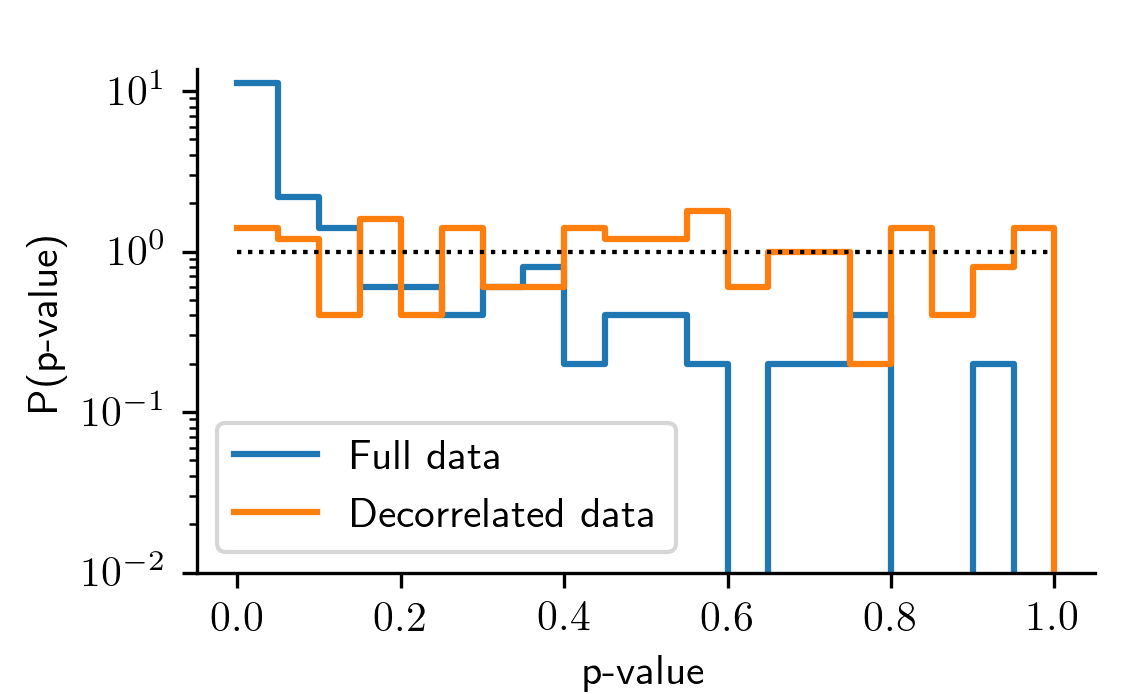}
\caption{
Synthetic data from a Gaussian distribution with long-range correlations~$C(\tau) \sim \tau^{-\gamma}$,  with $\gamma=0.1$..
The synthetic data consists of $N=2^{18}$ points generated as in Bunde et al., Phys. Rev. Lett. 94, 048701 (2005).
Left: Distribution $p(x)$ of the empirical data (blue) and the asymptotic Gaussian (orange).
Middle: Autocorrelation $C(\tau)$ of the empirical sequence of data (blue) and the shuffled sequence (orange).
Right: Distribution of p-values from 100 different synthetic datasets of size $N=2^{18}$ for the full correlated data (blue) and the decorrelated data by subsampling to $N^*=N/\tau^*$.
}
\label{fig.synth.lrc}
\end{figure}

\end{widetext}

\begin{thebibliography}{99.}%


\bibitem{Mitzenmacher2004} M. Mitzenmacher,
  {\it A Brief History of Generative Models for Power  Law and Lognormal Distributions},
  Internet Mathematics 1, 226 (2004).

\bibitem{Newman} M. E. J. Newman, {\it Power Laws, Pareto Distributions and Zipf’s law},
  Contemp. Phys. 46, 323 (2005).

\bibitem{Laws} E. G. Altmann and M. Gerlach, {\it Statistical laws in linguistics},
  Chap. in Creativity and Universality in Language, M. Degli Esposti, E. G. Altmann, F. Pachet (Eds)., Lecture Notes in Morphogenesis (Springer, 2016) 

  \bibitem{Zipf} G. K.  Zipf, {\it The Psycho-Biology of Language}  (Routledge,
  London, 1936). {\it Id.},  {\it Human behavior and the principle of least effort} (
  Addison-Wesley Press, Oxford, 1949).   

\bibitem{Sornette}  V. F. Pisarenko and D. Sornette, {\it Statistical Detection and Characterization of a Deviation from the Gutenberg-Richter Distribution above Magnitude 8},
  Pure and Applied Geophysics {\bf 161}, 839 (2004).
  
\bibitem{BarabasiAlbert1999} A. L. Barab\'asi and R. Albert, {\it Emergence of scaling in random networks},
  Science {\bf 286}, 509 (1999).

\bibitem{bursts} E. G. Altmann, J. B. Pierrehumbert, and A. E. Motter, {\it Beyond word frequency: Bursts, lulls, and scaling in the temporal distributions of words},
  PLoS ONE {\bf 4},  e7678 (2009).

\bibitem{burstsC} A. Corral, R. Ferrer-i-Cancho, G. Boleda, A. Diaz-Guilera, {\it Universal complex structures in written language},
  arXiv:0901.2924 (2009).

\bibitem{Bunde} A. Bunde, J. F. Eichner, J. W. Kantelhardt, S. Havlin, {\it Long-term memory: A natural mechanism for the clustering of extreme events and anomalous residual times in climate records},
  Phys. Rev. Lett. {\bf 94}, 048701 (2005).

\bibitem{Barabasi} A.-L. Barab\'asi, {\it The origin of burstiness and heavy tails in human dynamics},
  Nature {\bf 435}, 207 (2005). 

\bibitem{Broido} A. D. Broido and A. Clauset, {\it Scale-free networks are rare},
  Nature Communications {\bf 10}, 1017 (2019). 

\bibitem{Khanin} R. Khanin and E. Wit, {\it How Scale--Free are Biological Networks},
  Journal of Computational Biology {\bf 13}, 810 (2006).

\bibitem{Quanta2018} E. Klarreich, {\it Scant Evidence of Power Laws Found in Real-World Networks},
  Quanta Magazine, Feb. 15 (2018).

\bibitem{PRX} M. Gerlach and E. G. Altmann, {\it Stochastic Model for the Vocabulary Growth in Natural Languages},
  Phys. Rev. X 3, 021006 (2013).


\bibitem{Francesc} I. Moreno-S\'anchez, F. Font-Clos, A. Corral, {\it Large-Scale Analysis of Zipf’s Law in English Texts},
    PLOS ONE {\bf 11}, e0147073 (2016).


\bibitem{CriticalTruth} M. P. H. Stumpf and M. A. Porter, {\it Critical Truths About Power Laws},
  Science {\bf 335}, 665 (2012).


\bibitem{CSN} A. Clauset, C. R. Shalizi, and M. E. J. Newman, {\it Power-Law Distributions in Empirical Data},
  SIAM Review. 51, 661–703 (2009).
  
\bibitem{Goldstein04}
  M. L. Goldstein, S. A. Morris, and G. G. Yen. {\it Problems with fitting to the power-law distribution},
  Eur. J. Phys. B 41, 255-258 (2004). 

\bibitem{Bauke07}
  H. Bauke, {\it Parameter estimation for power-law distributions by maximum likelihood methods}
  Eur. J. Phys. B {\bf 58}, 167 (2007). 


\bibitem{Deluca13}
  A. Deluca and A. Corral, {\it Fitting and goodness-of-fit test of non-truncated and truncated power-law distributions},
  Acta Geophysica {\bf 61}, 1351 (2013).

  \bibitem{Hanel} R. Hanel, B. Corominas-Murtra, B. Liu, S. Thurner, {\it Fitting power-laws in empirical data with estimators that work for all exponents}, 
PLOS ONE {\bf 13}, e0196807 (2018).

\bibitem{StumpfEPL} M. P. H. Sumpf and P. J. Ingram, {\it Probability models for degree distributions of protein interaciton networks},
  EPL {\bf 71}, 152 (2005).


\bibitem{Eisler.2008} Z. Eisler, I. Bartos and J. Kert\'esz,  {\it Fluctuation scaling in complex systems: Taylor's law and beyond},
  Adv. Phys. {\bf 57}, 89 (2008).
  
\bibitem{StumpfPNAS} M. P. H. Stumpf, C. Wiuf, and R. M. May, {\it Subnets of scale-free networks are not scale-free: Sampling properties of networks},
  Proc. Nat. Acad. of Sci. USA {\bf 102}, 4221 (2005).


\bibitem{Lee} S. H. Lee, P.-J. Kim, and H. Jeong, {\it Statistical properties of sampled networks},
  Phys. Rev. E {\bf 73}, 016102 (2006).

\bibitem{StumpfPRE} M. P. H. Stumpf and C. Wiuf, {\it Sampling properties of random graphs: The degree distribution},
  Phys. Rev. E {\bf 72}, 036118 (2005).

\bibitem{Crane} H. Crane, {\it Probabilistic Foundations of Statistical Network Analysis}, Chapman and Hall/CRC, New York (2018).

\bibitem{B1} C.P. Robert, G. Casella, {\it Monte Carlo statistical methods}, Springer texts in statistics, 2nd ed. (Springer, Berlin, 2005). 

\bibitem{B2} D. A. Levin and Y. Peres, {\it Markov Chains and Mixing Times}, American Mathematical Society (2017).
  
 \bibitem{earthquakes} Southern California Earthquake Data Center \url{http://scedc.caltech.edu/research-tools/alt-2011-yang-hauksson-shearer.html} 

 \bibitem{Corral2004} A. Corral, {\it
    Long-term clustering, scaling, and universality in the temporal occurrence of earthquakes},
   Phys. Rev. Lett. {\bf 92}, 108501 (2004).

\bibitem{gutenberg} Project Gutenberg \url{http://www.gutenberg.org.} 

\bibitem{network} KONECT Project: Internet Topology \url{http://konect.cc/networks/topology} 

\bibitem{vankampen} N. G. Van Kampen, Stochastic Processes in Physics and Chemistry, (Elsevier, 1992).

\bibitem{Gasser} T. Gasser, {\it Goodness-of-fit tests for correlated data},
  Biometrika {\bf 51}, 563 (1975).
  
\bibitem{Weiss78}
  M. S. Weiss, {\it Modification of the Kolmogorov-Smirnov Statistic for Use with Correlated Data},
  J. Am. Stat. Assoc. {\bf 73}, 872 (1978).

\bibitem{Chicheportiche11}
  R. Chicheportiche and J.-P. Bouchaud, {\it Goodness-of-Fit tests with Dependent Observations},
  J. Stat. Mech., P09003 (2011).

\bibitem{bonferroni} J. P. Shaffer, {\it Multiple hypothesis testing},
  Annu. Rev. Psychol. {\bf 46}, 561 (1995).

\bibitem{dodds} P. S. Dodds, D. H. Rothman, and J. S. Weitz, {\it Re-examination of the ``3/4-law'' of Metabolism},
  J. Theor. Biol. {\bf 209}, 9 (2001).

\bibitem{us} J. C. Leitao, J. M. Miotto, M. Gerlach, and E. G. Altmann, {\it Is this scaling nonlinear?},
  R. Soc. Open Sci. {\bf 3}, 150649 (2016) .
  
\bibitem{Arratia2005} R. Arratia and T. M. Liggett, {\it How likely is an i.i.d. degree sequence to be graphical?},
  The Annals of Applied Probability {\bf 15}, 652 (2005).

\bibitem{Hastie09}
T. Hastie, R. Tibshirani, and J. Friedman,  {\it The elements of statistical learning}, Springer, New York  (2009).


\bibitem{Burnham02}
K. P. Burnham and D. R. Anderson, {\it Model selection and multimodal inference: a practical information-theoretic approach}, Springer, New York (2002).

\bibitem{Note1} Consider $x_i$'s to be a sequence of degrees of a network sampled from H1 and H2. For large networks, only half of the realizations lead to graphical degree sequences~\cite{Arratia2005}.

\bibitem{footnote-autocorrelation} For the data with assumed powerlaw distributions, we calculate the autocorrelation for the variable $\log x$. We determine $\tau^*$ as the minimum $\tau$ for which the lower bound $C(\tau)$ ($1$-percentile) of the original data is smaller or equal than the upper bound $C(\tau)$  ($99$-percentile) of the randomized data. We generate an ensemble of ``original'' timeseries of the same length $N$ by selecting a random point as the first observation and using periodic boundary conditions.

\bibitem{Note3} Assuming observations occur at an approximately constant rate over the 30 years of available data.
  
\bibitem{codes} Codes and datasets are available at: \url{https://doi.org/10.5281/zenodo.2641375}

  

  
\end{thebibliography}
\end{document}